\documentclass[5p]{elsarticle}
\usepackage[utf8]{inputenc}

\usepackage[colorlinks]{hyperref}
\hypersetup{
     colorlinks   = true,
     citecolor    = red,
	linkcolor=blue
}

\usepackage{bbm}
\usepackage{amsfonts}
\usepackage{mathrsfs}

\usepackage{amsmath,amssymb}

\usepackage{epsfig}
\usepackage{graphicx}               
\usepackage{url}
\usepackage{hyperref}
\usepackage{float}
\usepackage{soul}
\usepackage{pstricks}
\usepackage{color}
\usepackage{multirow}

\makeatletter
\def\simgt{\mathrel{\lower2.5pt\vbox{\lineskip=0pt\baselineskip=0pt
           \hbox{$>$}\hbox{$\sim$}}}}
\def\simlt{\mathrel{\lower2.5pt\vbox{\lineskip=0pt\baselineskip=0pt
           \hbox{$<$}\hbox{$\sim$}}}}
\makeatother

\newcommand{\be}{\begin{equation}}
\newcommand{\ee}{\end{equation}}
\newcommand{\bea}{\begin{eqnarray}}
\newcommand{\eea}{\end{eqnarray}}
\newcommand{\beq}{\begin{eqnarray}}
\newcommand{\eeq}{\end{eqnarray}}

\newcommand{\ellp}{l_p}

\def\lsim{\mathrel{\rlap{\lower4pt\hbox{\hskip1pt$\sim$}}
     \raise1pt\hbox{$<$}}}         
\def\gsim{\mathrel{\rlap{\lower4pt\hbox{\hskip1pt$\sim$}}
     \raise1pt\hbox{$>$}}}         

\begin{document}

\title{Observational Signatures of Quantum Gravity in Interferometers}
\author[Ams]{Erik P. Verlinde}
\author[Caltech]{Kathryn M. Zurek}

\address[Ams]{Institute of Physics, University of Amsterdam, Amsterdam, The Netherlands}
\address[Caltech]{Walter Burke Institute for Theoretical Physics, California Institute of Technology, Pasadena, CA USA}

\begin{abstract}
We consider the uncertainty in the arm length of an interferometer due to metric fluctuations from the quantum nature of gravity, proposing a concrete microscopic model of energy fluctuations in holographic degrees of freedom on the surface bounding a causally connected region of spacetime.  
In our model, fluctuations longitudinal to the beam direction accumulate in the infrared and feature strong long distance correlation in the transverse direction. This leads to a signal that could be observed in a gravitational wave interferometer.  We connect the positional uncertainty principle arising from our calculations to the 't Hooft gravitational $S$-matrix.  

\end{abstract}

\maketitle

\section{Introduction}
\label{sec:intro}
The quantum mechanical description of gravity together with the other forces remains one of the most important questions in physics.  While general relativity can be quantized as an effective field theory valid at low energies, and there has been significant {\em theoretical} progress in understanding other aspects of quantum gravity, signatures of the quantum nature of gravity have so far remained stubbornly immune to observation.  

An important clue towards the ultimate theory of quantum gravity is provided by the holographic principle \cite{tHooft:1993dmi,Susskind:1994vu}. One of its implications is the covariant entropy bound  \cite{Bousso:1999xy}, which states that the entropy associated to region bounded by null geodesics 
does not exceed $A/4 G_N$, where $A$ denotes the area of the surface and Newton's constant is identified with the square of the Planck length  via $8\pi G_N \equiv \ellp^2$, with $\ellp \simeq 10^{-35} \mbox {m}$. The holographic principle suggests that the total number of microscopic degrees of freedom associated to a given region of space (defined by the maximal entropy) is given by the area of the surrounding surface in Planck units. In this form the holographic principle is known to be realized in spacetimes with negative cosmological constant \cite{Susskind:1998dq}, and is firmly incorporated in the framework of the AdS/CFT correspondence \cite{Maldacena:1997re}.

Motivated by the holographic principle, one is tempted to postulate that the microscopic spacetime degrees of freedom, also in flat spacetime, can be identified with Planck size pixels on the surface bounding a causally connected part of space. The spacetime volume would then emerge in the infrared from these holographic spacetime quanta.  
One of the intriguing aspects of the holographic principle is that, to ensure the validity of the entropy bound,  the spacetime degrees of freedom are necessarily {\em correlated} in the infrared.   This raises the question of whether Planck scale physics could appear at much longer, potentially observable, length scales.  

Our goal in this {\em Letter} is to investigate whether fluctuations due to the graininess of spacetime  can potentially lead to observable signatures. Normal intuition would say that, since the natural length and time scale associated with the quantum nature of spacetime is Planckian, that no feasible experiment exists that could measure its effects. We will argue, however, that when combined with important infrared effects naturally expected from holography, the accumulative effect of Planck scale fluctuations can be transmuted to observable time and length scales. Because of their sensitivity to exquisitely short distance scales, gravitational wave interferometers are an ideal testing ground for these ideas. 

We will identify the required theoretical conditions that need to be satisfied to obtain observable effects, and construct a concrete holographic model that obeys those conditions. After showing that uncorrelated Planckian fluctuations are not macroscopically observable, we demonstrate that fluctuations with sufficient transverse correlations in the infrared do lead to observable effects.  The appearance of transverse correlations is crucial and suggests a holographic description in which the longitudinal and transverse behavior of the spacetime degrees of freedom  are treated on a different footing.  

We will build an explicit holographic model   in terms of Planck-size pixels that saturate the holographic bound 
     and have energy fluctuations that cause the spatial length $L$ of a causally  connected region of space to fluctuate.  The transverse correlations are generated through the Newtonian potential of these energy fluctuations, allowing us to make a concrete prediction for the spectrum of length fluctuations in an interferometer.  These fluctuations imply a spacetime uncertainty relation in the longitudinal direction, which we connect, albeit in a modified form, to the gravitational $S$-matrix approach of 't Hooft (see Refs.~\cite{tHooft:1996rdg,Hooft:2018syc}).

  While there have been several previous studies seeking to heuristically connect holography to interferometry ({\em e.g.} \cite{Ng:1993jb,Ng:1995km, AmelinoCamelia:1998ax, Hogan:2007pk, Kwon:2014yea}), our theoretical description is structurally unique in its holographic set-up.  And although the first steps we take--employing a Planckian random walk--shares commonalities with these works, our approach differs in the sense that we present a concrete theoretical model leading to length fluctuations along the longitudinal direction with a distinctive signature for strong transverse correlations, which is, as a consequence, macroscopically observable in an interferometer.  A phenomenological result is that constraints from the images of distant astrophysical sources derived for uncorrelated fluctuations in Refs. \cite{Perlman:2011wv,Perlman:2014cwa} do not apply to our model.   
 
\section{Length Fluctuations with Planckian White Noise}  
In this paper we consider a toy experimental set-up, shown in Fig.~\ref{fig:interferometer}, in which the arm length $L$ of an interferometer is measured after a single light crossing. In this idealized scenario the length fluctuations $\delta L$ due to quantum fluctuations in the metric is given by\\[-5mm]  
 \beq
\label{eq:strain-integral}
\delta L(t) = \frac{1}{2}\int_0^L\!\!dz\, h(t\!+\!z\!-\!L) 
\eeq
where $h\equiv h_{zz}$ is the metric component along the light beam propagation (see {\em e.g.} \cite{LIGOTN}).  The magnitude of these length fluctuations is normally expressed in terms of the power spectral density (PSD)
 \beq
S(\omega,t) = \int_{-\infty}^\infty\!\! d\tau \left\langle \frac{\delta L(t)}{L}\frac{\delta L(t-\tau)}{L}  \right\rangle e^{-i\omega \tau}.
\label{eq:PSD}
 \eeq  
Let us first consider a simple model with a white noise signal of Planckian amplitude
\beq
 \bigl \langle  h(t\!+\! z_1 \!-\! L) h(t\!+\!z_2\!-\! L\!-\!\tau)  \bigr\rangle = C \ellp \delta(\tau \!+\!z_1\!- \!z_2),
 \label{eq:PlanckNoise}
 \eeq 
 where 
 $\ellp = \sqrt{8 \pi G_N}$. 
This leads to a PSD of the form
\beq
   S(\omega) = {C\ellp \over 4} {\sin^2 \omega L\over \omega^2 L^2}. 
      \label{eq:SfiltdeltaL}
   \eeq
In this simple model the length fluctuations $\langle \delta L^2\rangle $  obey 
\beq
\left\langle{ \delta L^2 (t)\over L^2}\right\rangle = {1\over 2\pi}\int_{-\infty}^{\infty}\! \!\! d\omega \, S(\omega)\,= {C \ellp \over 8 L},
 \label{eq:RMSflucs}
 \eeq
 and thus grow linearly with $L$ \cite{Ng:1993jb,Ng:1995km, AmelinoCamelia:1998ax, Hogan:2007pk, Kwon:2014yea}. This signal could in principle be observable, since the peak sensitivity for gravitational wave interferometers is right around the Planck scale: $S(\omega,t) \lesssim \ellp$.  Over the next sections our goal will be to show how some of the generic behavior in Eqs.~\ref{eq:SfiltdeltaL},~\ref{eq:RMSflucs} can arise from a holographic model, motivating the size of the constant $C$, with crucial observational effects arising from angular correlations. 
 In addition, in experiments like LIGO and Virgo a typical photon traverses the interferometer arm multiple times before being measured. In this paper we continue to focus on our simple set up and defer the detailed discussion of multiple crossings to future work.

 \section{Holographic Scenario and Basic Postulates}
\begin{figure}[t]
\begin{center}
\vspace{-0.2cm}
\includegraphics[scale=0.37]{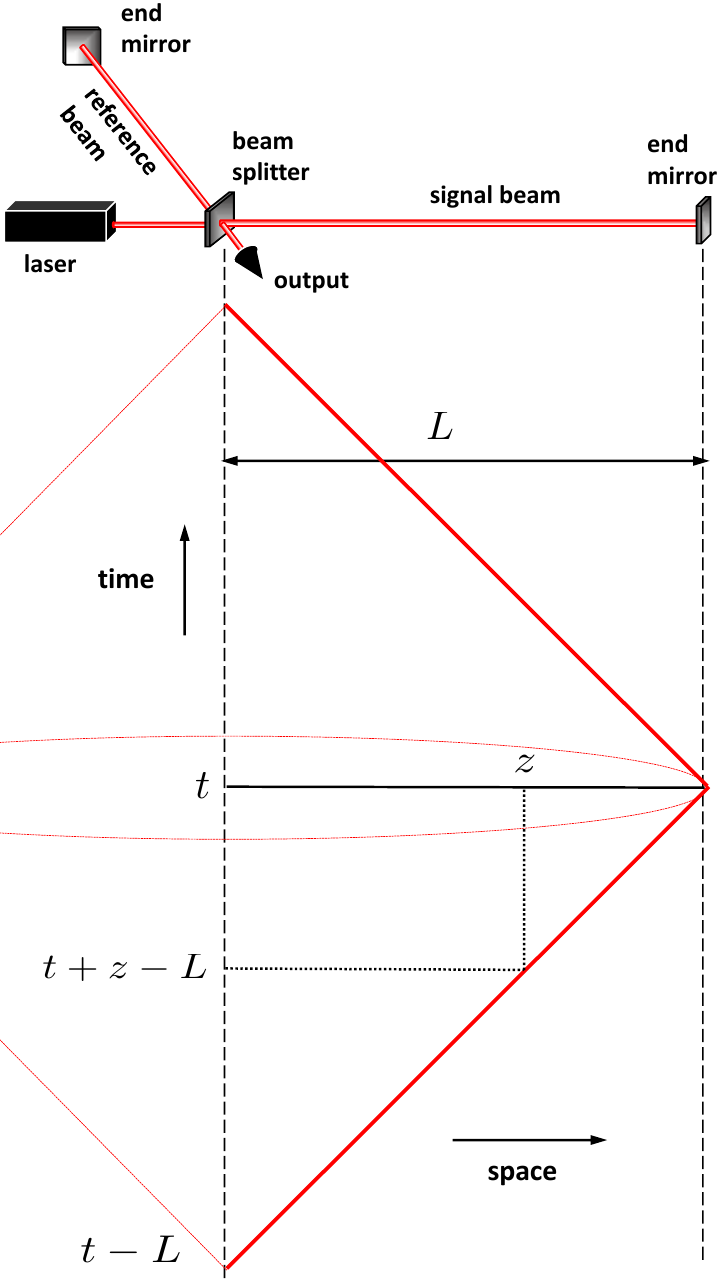}
\vspace{-0.2cm}
\caption{The interferometer together with the spacetime diagram for a single crossing of a photon in the signal beam.  The interferometer at time $t$ is contained in a causal diamond centered at the beam splitter and with the photon path on its null boundary. \label{fig:interferometer}}
\end{center}
\end{figure}

 Our aim in the following is to derive a result similar to Eq.  \ref{eq:RMSflucs} from a holographic scenario, in which 
  the holographic surface is fixed by the light path of a photon, as depicted in Fig.~\ref{fig:interferometer}. In order to clearly
delineate between theoretical input and observational consequences, we will state here our basic postulates:
\begin{enumerate}

\item { \em Holographic principle in flat spacetime.}  We postulate that the holographic principle also applies to Minkowski spacetime. It states that the maximal entropy carried by the microscopic degrees of freedom associated with a finite region of flat spacetime bounded by null geodesics is $S= A/4G_N$. This bound is saturated for a region of space whose null boundary coincides with a horizon.
 
 \item{ \em Universality of metric fluctuations at horizons.}
We postulate, as a corollary of the first postulate, that metric fluctuations near null surfaces associated with the boundary of a finite region follow from the entropy and temperature using standard thermodynamic considerations. This postulate implies that metric fluctuations near a Rindler-type horizon are identical to those near a black hole horizon with the same temperature and entropy.

\end{enumerate} 

Note that we are treating the metric fluctuations at the holographic surface separating the inside of the causal diamond from the outside as if it were a black hole horizon (see Ref.~\cite{Marolf:2003bb}), even though we are considering the vacuum of Minkowski space.  The basic reason we believe these are reasonable postulates is that a finite causal diamond in Minkowski space, when suitably foliated, can be recast in the metric of a so-called topological black hole \cite{CHM}.  Furthermore, a conformal field theory restricted to the diamond behaves as a thermal field theory \cite{CHM}, and the quantized Einstein-Hilbert metric in the infrared behaves as a conformal field theory.  In related work \cite{VZ2}, we show that these postulates are justified in the context of AdS/CFT.  That they hold for the Einstein-Hilbert metric in Minkowski space must, at the present time, be ultimately verified by experiment.  Fortunately, we show that the experimental signatures associated with a spacetime obeying these postulates are within reach with current interferometer technology. 

\section{Towards Macroscopic Effects in Interferometers}
The results in Eqs.~\ref{eq:PlanckNoise}-\ref{eq:RMSflucs}, that were derived from the simple 1D-model, are by themselves not sufficient to show an effect. In order to be observable in a realistic experimental set up, the fluctuations must be coherent at macroscopic spacetime distances. To examine the conditions under which such coherent fluctuations occur, we extend our model by including the two spatial directions transverse to the beam direction. Anticipating our holographic description, we consider metric fluctuations that depend on only three coordinates, one longitudinal null direction and two transversal directions, corresponding to the outside boundary of the causal diamond in Fig.~\ref{fig:interferometer},  
\beq
 \left\langle \left(\frac{\delta L}{L}\right)_{\!1}\!\left(\frac{\delta L}{L}\right)_{\!2} \right\rangle 
    &= &    \frac{1}{16 L^2}  \int_{-L}^L \int_{-L}^L du_1 du_2   \\ \nonumber    \int  \frac{d^3 k_1}{(2\pi)^3}  \frac{d^3 k_2}{(2\pi)^3}   & &  \langle  h(k_1)   h(k_2)  \rangle  e^{i k_1 \cdot x_1} e^{i k_2 \cdot x_2}. 
  \label{eq:RMSFourierLightCone}
\eeq
We first consider an Ansatz corresponding to uncorrelated white noise in those three dimensions:
\beq
\bigl\langle h(k_1) h(k_2)\bigr\rangle = (2\pi)^3 \delta^3(k_1 + k_2)  C \ellp^3.
\label{eq:3dPlanckPower}
\eeq
This power spectrum implements the principle of statistical independence both in the longitudinal as well as the transversal directions.  This can be seen directly by computing the PSD and RMS length fluctuations: 
 \beq
   \label{eq:3dFiltPSD}
  \left\langle \left(\frac{\delta L}{L}\right)_{\!1}\!\left(\frac{\delta L}{L}\right)_{\!2} \right\rangle & = & {C\ellp\over 16\pi  L} { 1\over (\Delta x_T^2 /\ellp^2  + 1)^{3/2}}. 
  \eeq
   In the limit $\Delta x_T \rightarrow 0$, we recover a signal of an amplitude that is in principle within the observable range and is consistent with Eqs.~\ref{eq:SfiltdeltaL}-\ref{eq:RMSflucs}.  However, for a realistic macroscopic interferometer, with the beam size centimeters across such that $\Delta x_T/\ellp \!\gg\! 1$, this signature would be unobservable.  
   
   Let us consider an alternative Ansatz for the metric fluctuations in which the transversal directions are treated differently:
  \beq
\bigl\langle h(k_1) h(k_2)\bigr\rangle = (2\pi)^3 \delta^3(k_1 + k_2)  \frac{C\ellp}{(k_T^2+k_{IR}^2)},
\label{eq:TransversePower}
\eeq
where $k_{IR}$ acts as a regulator.  Then Eq.~(\ref{eq:3dFiltPSD}), in the limit that $k_{\rm IR} \Delta x_T \ll 1$, becomes 
   \beq
  \left\langle \left(\frac{\delta L}{L}\right)_{\!1}\!\left(\frac{\delta L}{L}\right)_{\!2} \right\rangle & \sim &  {C\ellp\over 16 \pi  L}  \log\left[1/\Delta x_T k_{IR}\right].
   \label{eq:TransverseFiltPSD}
   \eeq

Already this result shows important features that the underlying theory must give, notably that the longitudinal and transverse directions appear on a different footing.  The metric fluctuation in the transverse direction must be correlated, while the metric fluctuations in the longitudinal direction accumulate, as in a random walk, and are transmuted to a low-energy, long-distance signature.
 We will show over the next sections how these features arise naturally from energy fluctuations on a holographic surface.      

\section{From Minkowski to Schwarzschild-like Metric}
The central part of our argument involves utilizing a correspondence between any horizon and a black hole horizon.  To show concretely how this applies to the case at hand, we make two metric transformations, which are described below.  First we define light cone coordinates $u = r+t$ and $v = r-t $ so that metric becomes
\beq
\label{eq:fluctuating-metric}
ds^2 = du dv+ dy^2 + h_{uu} du^2 + h_{vv} dv^2 + \ldots
\eeq
where the dots denote the angular components.   In this metric the light paths on the lower and upper half of the causal diamond shown in Fig.~\ref{fig:interferometer} are given by
\vspace{0mm}
$$
v= L +\delta v(u)  \qquad \mbox{ and }\qquad u = L+\delta u(v)\\[-0mm]
$$
The total length fluctuation $\delta L$ can be expressed as
\vspace{0mm}
$$
\delta L = \left(\delta v(L) + \delta u(L)\right)/2.\\[-1mm]
$$
It turns out that only one metric component  contributes to the time delay along each light path.  As we will show in a companion paper, the values for $\delta v(L)$ and $\delta u(L)$ can be expressed in terms of the metric fluctuations via 
\vspace{0mm}
\begin{eqnarray}
\delta v(L) & = & 
\!\int_{-L}^L \! \!\! du \  h_{uu} (u,L) 
\\ \nonumber
\delta u(L)  & = & \int_{-L}^{L} \! \!\! dv \ h_{vv} (L,v). \label{eq:deludelv}
\end{eqnarray} As a next step, to employ our postulates,  we recast the metric in the Schwarzschild-like form \vspace{0mm}
\beq
\label{eqn:schwarzschild}
ds^2 = -f(R) dT^2 + {dR^2 \over f(R)} + r^2 (d\theta^2 +\sin^2\theta d\phi^2).
\eeq
in such a way that the light paths of the photon are mapped onto the event horizon located at $f(R)=0$. 
This is achieved by making the coordinate transformation
\vspace{0mm}
\beq
\mbox{${}$}\!\!\! (u-L)(v-L) = 4L^2 f(R),\qquad  
\log {u-L\over v-L} = {T\over L}\quad
\eeq
where the function $f(R)$ is given by
\vspace{0mm}
\beq
f(R) = 1- {R\over L} + 2 \Phi. 
\eeq
Here $\Phi$ plays the role of the Newtonian potential and parametrizes the deviations in the geometry due to vacuum fluctuations in the energy conjugate to the time $T$.  

Without any quantum gravity effects, the horizon is located at $R=L$. In general, its location is determined by $f(R)=0$. This leads to the following relationship between the product of the lightcone time variations $\delta u(L)$ and $\delta v(L)$ and the value of Newton's potential 
\beq
{\delta v(L) \delta u(L)\over L^2}= 2 \Phi(L).
\label{eq:NewtPot}
\eeq
This equation should be regarded as an operator identity.
Since $\langle \Phi \rangle = 0$ in vacuum,  the right-hand-side of this equation actually represents a fluctuation around the vacuum, whose amplitude is given by squaring the operators and taking its expectation value: 
\beq %
\left\langle\left({\delta v(L) \delta u(L)\over L^2}\right)^2\right\rangle =  \Bigl\langle 4 \Phi(L)^2 \Bigr\rangle .
\label{eq:fourpoint}
\eeq
For a more detailed and formal discussion of this point in the context of AdS/CFT, we encourage the reader to consult Sec IV of our companion paper Ref.~\cite{VZ2}.

The goal of the next section is to determine the root-mean-square value of the fluctuations in $\Phi$ in an ensemble averaged over many interferometer light crossings. 

\section{Holographic Model for Spacetime Fluctuations}

We are now ready to employ all of our postulates together to compute the deviations in the Newtonian potential, Eq.~\ref{eq:NewtPot}.  The fluctuations in $\Phi(L)$ will be induced by vacuum fluctuations in the energy conjugate to the time coordinate $T$.  Eq.~\ref{eq:NewtPot}. In the following analysis we follow closely the reasoning of Marolf in Ref.~\cite{Marolf:2003bb} for the quantum thickness of black hole horizons. Directly applying the holographic principle to the horizon of the causal diamond gives
\beq
S_{hor} = \frac{A}{4 G_N} = \frac{8 \pi^2 L^2}{\ellp^2}.
\label{eq:ent}
\eeq
Now the fluctuations in the Newtonian potential on the horizon obeys
\beq
2\Phi(L) = -\frac{\ellp^2\Delta M}{4\pi L},
\eeq
where $\Delta M$ represents the energy fluctuations in the holographic degrees of freedom. Heuristically, one expects the RMS value of $\Delta M$  to scale as the square root of the number of pixels on the horizon, times the typical energy of the fluctuation, which is given by the Hawking temperature. 

One of the standard methods to determine the Hawking temperature is to go to Euclidean time and impose that the resulting metric is free from conical singularities. In this way one finds 
\beq
T_{hor} = {|f'(L)|\over 4\pi}= \frac{1}{4\pi L}.
\eeq
In the present situation the temperature $T_{hor}$ is measured by an accelerated observer whose event horizon coincides with the photon trajectory and whose own trajectory passes through the origin at $T=0$. This observer stays at $R=0$ and has $T$ as proper time coordinate.  

We now calculate the RMS value of the fluctuations, by assuming that the vacuum energy $E$ vanishes. This implies that the free energy $F(\beta)$ equals 
\beq
F(\beta)=-T_{hor} S_{hor} = -{\beta\over 2  l_p^2}
\eeq
where in the last step we eliminated the length $L$ in favor of the inverse temperature $\beta = 1/T_{hor}=4\pi L$. In the canonical ensemble the mass fluctuations $\Delta M$ are obtained by taking the second derivative of the free energy. One thus obtains
\beq
\langle \Delta M^2\rangle =-{\partial^2\over \partial \beta^2} \left(\beta F\right) ={1\over l_p^2}.
\label{eq:DelM}
\eeq
Note that $\Delta M\sim  T_{hor}\sqrt{S_{hor}}$, as expected from the heuristic argument.
We now assume that at a coincident point, $\delta v(L)$ and $\delta u(L)$ take the same value $\delta L$. In this situation, combining Eqs.~\ref{eq:NewtPot}-\ref{eq:DelM}, we learn that the amplitude of the length fluctuation is
\beq
\left\langle {\delta L^2\over L^2} \right\rangle =  {{\ellp^2}  \Delta M \over 4\pi  L} =  {\ellp \over 4\pi L},
\label{eq:deltaLsq}
\eeq
where here $\Delta M = \sqrt{\langle \Delta M^2 \rangle}$ is interpreted as the root-mean-square of the mass fluctuation.  Note this has precisely the behavior shown in Eq.~\ref{eq:RMSflucs} needed to be observable, where now we can fix the constant $C$ via the holographic principle.  
We will propose in the next section that angular correlations between the interferometer arms respect the spherical symmetry of the measuring apparatus, and would give rise to a distinctive experimental signature.

\section{Angular Correlations and 't Hooft's $S$-matrix}
We have considered so far the amplitude of the fluctuations only as a function of the longitudinal coordinates. Physically it is clear that the fluctuations will also have an angular dependence, which can be straightforwardly determined for an interferometer with two arms of equal length $L$.  In this case, a spherical coordinate system, with origin at the beamsplitter, is appropriate, with the far mirrors located at two positions $\tilde {\bf r}_1,~\tilde {\bf r}_2$ on the sphere.  In this experimental configuration, the angular information can be determined with the help of the Newtonian potential $\Phi$ decomposed in terms of spherical harmonics, thus respecting the spherical symmetry of the measuring apparatus.  

Accordingly, we propose the following natural Ansatz as a generalization of Eq.~\ref{eq:deltaLsq}: 
\beq
{}\!\!\!\Bigl \langle \delta L(\tilde{\bf r}_1) \delta L(\tilde{\bf r}_2) \Bigr\rangle
=  {\ellp L\over 4\pi} \, {\bf G}(\tilde{\bf r}_1, \tilde{\bf r}_2). 
\label{eq:<hh2>}
\eeq 
This equation can be further justified from a generalization of Eq.~\ref{eq:fourpoint} to give the four-point correlation at separated points:
\vspace{0mm}
\beq
{}\!\!\!\Bigl \langle \delta u(\tilde{\bf r}_1) \delta v(\tilde{\bf r}_1) \delta u(\tilde{\bf r}_2) \delta v(\tilde{\bf r}_2) \Bigr\rangle
=  {\ellp^2 L^2\over 16\pi^2} \, {\bf G}^2(\tilde{\bf r}_1, \tilde{\bf r}_2). 
\label{eq:<hh>}
\eeq
If we again assume as above that at a coincident point $\delta u(\tilde{\bf r}) = \delta v(\tilde{\bf r}) = \delta L(\tilde{\bf r})$, the two-point Eq.~\ref{eq:<hh2>} is seen to be the factorization of Eq.~\ref{eq:<hh>} into the product of two two-point functions. 

Our suggestion, based on the symmetries of the system, is to identify the function  ${\bf G}(\tilde{\bf r}_1, \tilde{\bf r}_2)$ with the Green function of a modified Laplacian on the sphere. It obeys 
\beq
\left(-\nabla^2_{\tilde{\bf r}_1}+{1\over L^2}\right) {\bf G}(\tilde{\bf r}_1, \tilde{\bf r}_2)= \delta^{(2)}(\tilde{\bf r}_1, \tilde{\bf r}_2),
\eeq
and can be obtained by integrating the 3D Green function along the radial direction corresponding to the beam. 
At short distances it behaves as the normal Green function on the 2D-plane
\vspace{0mm}
\beq
\mbox{${}$}\!\!\!\!{\bf G}(\tilde{\bf r}_1, \tilde{\bf r}_2)  \sim {1\over 2\pi} \log\left({L\over |\tilde{\bf r}_1\!-\!\tilde{\bf r}_2|}\right)\quad \mbox{for $|\tilde{\bf r}_1\!-\!\tilde{\bf r}_2|\!<\!
\!<\! L$}.
\label{eq:GF}
\eeq
In terms of spherical harmonics it has the expansion
\vspace{0mm}
\beq
{\bf G}(\tilde{\bf r}_1, \tilde{\bf r}_2)  = \sum_{\ell,m} {Y_{\ell,m}(\tilde{\bf r}_1) Y^*_{\ell,m}(\tilde{\bf r}_2) \over \ell^2+\ell+1}.\label{eq:GFexpand}.
\eeq  
We can write this result alternatively
in terms of the coefficients, $\delta L_{\ell m}$ and $\delta_{\ell' m'}$, of the decomposition of $\delta L(\tilde{\bf r})$ and $\delta L(\tilde{\bf r})$ in to spherical harmonics as
\vspace{0mm}
\beq
\bigl\langle \delta L_{\ell m} \delta L_{\ell' m'}\bigr\rangle = {1\over 4\pi} {{\ellp L} \over {\ell^2+\ell+1}}\delta_{\ell \ell'}\delta_{m m'}.
\label{eq:sphharmunc}
\eeq
This relation tells us that much of the power in the fluctuations is contained in the low $\ell$ modes, and thus appears on the {\em largest} scales, contrary to one's intuition about Planckian effects.  While we think that our Ansatz of fluctuations obeying the Green function on the 2-d sphere is natural given the symmetries of the system, we have not rigorously derived this result here; we leave more detailed work in this direction for the future.

Our result implies a fundamental uncertainty relation between the longitudinal spacetime components. 
As 't Hooft showed, the in-going and out-going radiation at the horizon causes a spacetime shift due to gravitational shock waves. 
 He then went on to postulate that there is an inherent uncertainty in the values of the position of the horizon. In fact, 't Hooft's uncertainty relations described in {\em e.g.} \cite{tHooft:1996rdg,Hooft:2018syc}, when translated into a correlation function, have an identical angular dependence as Eq.~\ref{eq:sphharmunc}, though a different normalization. Due to our assumption of statistical independence and the resulting accumulation of the spacetime fluctuations, we find an extra factor $L/\ellp$ compared to 't Hooft.

\section{Conclusion and Discussion}
In this {\em Letter} we have constructed a concrete holographic model of transversally correlated longitudinal distance fluctuations, due to vacuum energy fluctuations of the (holographic) degrees of freedom associated with a causally connected volume of spacetime. 
By assuming that the energy of these fluctuations is of the order $1/L$, where $L$ is the length of the interferometer arm, and that number of fluctuating degrees of freedom is $L^2/\ellp^2$, we have derived length fluctuations of size $\delta L^2 \sim \ellp L$. The strong transverse correlation implies that in interferometer experiments the length fluctuations are sufficiently coherent across the light beam, so that they are in principle observable. 

If a signal with the characteristic features of our model is observed it would be a confirmation that our postulates regarding the theory of quantum gravity in flat spacetime are realized.  Conversely, if no signal is observed in an appropriately sensitive interferometer, it would tell us that one our proposed postulates does not hold.  In either case, we will have obtained concrete experimental information about the underlying theory of quantum gravity.

Our results were derived for a simple toy Michelson interferometer, so the next step is to concretely connect the result in Eq.~\ref{eq:sphharmunc} to a power spectral density and to realistic interferometers.  The closest experimental set-up to our toy is the ``Holometer'' \cite{Chou:2017zpk}, but to make a concrete comparison to experimental results requires at minimum a computation extending Eq.~\ref{eq:sphharmunc} to include the (likely $\ell$-dependent) frequency information in the PSD.  
Gravitational wave interferometers like LIGO and Virgo have multiple light-crossings; in our model we expect the signal from each subsequent light-crossing to be statistically uncorrelated, effectively reducing the signal in Eq.~\ref{eq:sphharmunc} by a factor of the number of light crossings (as discussed in Ref.~\cite{Kwon:2014yea}). To fully determine the observational implications of our results one needs to incorporate these and other experimental aspects in to our model. 

The transversal correlation of our model also has other important phenomenological implications. Previous attempts to consider phenomenological effects from Planckian Brownian noise (see Refs.~\cite{Christiansen:2009bz,Perlman:2011wv}, and as suggested by Eq.~\ref{eq:PlanckNoise}-\ref{eq:RMSflucs}) were stymied by the blurring of images from distant astrophysical sources. If the fluctuations are uncorrelated in the transverse direction, this implies large deviations in the phase of the light \cite{Perlman:2014cwa}. In the model under consideration here, however, most of the power in the length fluctuations is contained in low $\ell$-modes, as shown in Eq.~\ref{eq:sphharmunc}, that are coherent across the aperture diameter $D$ of an optical device, namely those with $\ell \leq 2\pi L/D$. These modes do not affect the quality of the image of astronomical objects. In our model the image quality is therefore improved compared to previously considered situations.  To fully establish that the transversal correlations are sufficient to evade the astrophysical constraints needs further study. 

On the theoretical side the most important open question is the precise nature of the holographic degrees of freedom that are responsible for the spacetime fluctuations. A possible route to gain more control over the microscopic theory is to consider an interferometer in Anti-de~Sitter space, and to reformulate the problem in terms of observables of the dual conformal field theory. The question is then to isolate the microscopic degrees of freedom dual to a small causal diamond deep inside the AdS geometry.  Another promising direction is to relate our description to the recent works on the BMS-group, soft gravitons and gravitational memory effects \cite{He:2014laa,Strominger:2014pwa}. In this context one is dealing with coordinate shifts at spatial infinity, or at event horizons of black holes \cite{Hawking:2016msc}. The theoretical challenge in this case is to generalize these studies to finite size causal diamonds, and determine the fluctuation spectrum of the coordinate shifts. We will leave these theoretical, as well as the phenomenological, analyses to future work.

{\em Acknowledgments.}
We thank Rana Adhikari for discussions on the experimental prospects, Cliff Cheung for discussions and comments on the draft, as well as Craig Hogan and Grant Remmen for conversation at the very early stages of this work.  KZ thanks the CERN theory group for hospitality over much of the duration of this work. 
The research of EV is supported by the grant ``Scanning New Horizons" and a Spinoza grant of the Dutch Science Foundation (NWO). 

\bibliography{QG}

\begin{thebibliography}{10}
\expandafter\ifx\csname url\endcsname\relax
  \def\url#1{\texttt{#1}}\fi
\expandafter\ifx\csname urlprefix\endcsname\relax\def\urlprefix{URL }\fi
\expandafter\ifx\csname href\endcsname\relax
  \def\href#1#2{#2} \def\path#1{#1}\fi

\bibitem{tHooft:1993dmi}
G.~'t~Hooft, {Dimensional reduction in quantum gravity}, Conf. Proc. C930308
  (1993) 284--296.
\newblock \href {http://arxiv.org/abs/gr-qc/9310026}
  {\path{arXiv:gr-qc/9310026}}.

\bibitem{Susskind:1994vu}
L.~Susskind, {The World as a hologram}, J. Math. Phys. 36 (1995) 6377--6396.
\newblock \href {http://arxiv.org/abs/hep-th/9409089}
  {\path{arXiv:hep-th/9409089}}, \href {https://doi.org/10.1063/1.531249}
  {\path{doi:10.1063/1.531249}}.

\bibitem{Bousso:1999xy}
R.~Bousso, {A Covariant entropy conjecture}, JHEP 07 (1999) 004.
\newblock \href {http://arxiv.org/abs/hep-th/9905177}
  {\path{arXiv:hep-th/9905177}}, \href
  {https://doi.org/10.1088/1126-6708/1999/07/004}
  {\path{doi:10.1088/1126-6708/1999/07/004}}.

\bibitem{Susskind:1998dq}
L.~Susskind, E.~Witten, {The Holographic bound in anti-de Sitter space} (1998).
\newblock \href {http://arxiv.org/abs/hep-th/9805114}
  {\path{arXiv:hep-th/9805114}}.

\bibitem{Maldacena:1997re}
J.~M. Maldacena, {The Large N limit of superconformal field theories and
  supergravity}, Int. J. Theor. Phys. 38 (1999) 1113--1133, [Adv. Theor. Math.
  Phys.2,231(1998)].
\newblock \href {http://arxiv.org/abs/hep-th/9711200}
  {\path{arXiv:hep-th/9711200}}, \href
  {https://doi.org/10.1023/A:1026654312961, 10.4310/ATMP.1998.v2.n2.a1}
  {\path{doi:10.1023/A:1026654312961, 10.4310/ATMP.1998.v2.n2.a1}}.

\bibitem{tHooft:1996rdg}
G.~'t~Hooft, {The Scattering matrix approach for the quantum black hole: An
  Overview}, Int. J. Mod. Phys. A11 (1996) 4623--4688.
\newblock \href {http://arxiv.org/abs/gr-qc/9607022}
  {\path{arXiv:gr-qc/9607022}}, \href
  {https://doi.org/10.1142/S0217751X96002145}
  {\path{doi:10.1142/S0217751X96002145}}.

\bibitem{Hooft:2018syc}
G.~'t~Hooft, {Discreteness of Black Hole Microstates} (9 2018).
\newblock \href {http://arxiv.org/abs/1809.05367} {\path{arXiv:1809.05367}}.

\bibitem{Ng:1993jb}
Y.~J. Ng, H.~Van~Dam, {Limit to space-time measurement}, Mod. Phys. Lett. A9
  (1994) 335--340.
\newblock \href {https://doi.org/10.1142/S0217732394000356}
  {\path{doi:10.1142/S0217732394000356}}.

\bibitem{Ng:1995km}
Y.~J. Ng, H.~Van~Dam, {Remarks on gravitational sources}, Mod. Phys. Lett. A10
  (1995) 2801--2808.
\newblock \href {https://doi.org/10.1142/S0217732395002945}
  {\path{doi:10.1142/S0217732395002945}}.

\bibitem{AmelinoCamelia:1998ax}
G.~Amelino-Camelia, {An Interferometric gravitational wave detector as a
  quantum gravity apparatus}, Nature 398 (1999) 216--218.
\newblock \href {http://arxiv.org/abs/gr-qc/9808029}
  {\path{arXiv:gr-qc/9808029}}, \href {https://doi.org/10.1038/18377}
  {\path{doi:10.1038/18377}}.

\bibitem{Hogan:2007pk}
C.~J. Hogan, {Measurement of Quantum Fluctuations in Geometry}, Phys. Rev. D77
  (2008) 104031.
\newblock \href {http://arxiv.org/abs/0712.3419} {\path{arXiv:0712.3419}},
  \href {https://doi.org/10.1103/PhysRevD.77.104031}
  {\path{doi:10.1103/PhysRevD.77.104031}}.

\bibitem{Kwon:2014yea}
O.~Kwon, C.~J. Hogan, {Interferometric Tests of Planckian Quantum Geometry
  Models}, Class. Quant. Grav. 33~(10) (2016) 105004.
\newblock \href {http://arxiv.org/abs/1410.8197} {\path{arXiv:1410.8197}},
  \href {https://doi.org/10.1088/0264-9381/33/10/105004}
  {\path{doi:10.1088/0264-9381/33/10/105004}}.

\bibitem{Perlman:2011wv}
E.~S. Perlman, Y.~J. Ng, D.~J.~E. Floyd, W.~A. Christiansen, {Using
  Observations of Distant Quasars to Constrain Quantum Gravity}, Astron.
  Astrophys. 535 (2011) L9.
\newblock \href {http://arxiv.org/abs/1110.4986} {\path{arXiv:1110.4986}},
  \href {https://doi.org/10.1051/0004-6361/201118319}
  {\path{doi:10.1051/0004-6361/201118319}}.

\bibitem{Perlman:2014cwa}
E.~S. Perlman, S.~A. Rappaport, W.~A. Christiansen, Y.~J. Ng, J.~DeVore,
  D.~Pooley, {New Constraints on Quantum Gravity from X-ray and Gamma-Ray
  Observations}, Astrophys. J. 805~(1) (2015) 10.
\newblock \href {http://arxiv.org/abs/1411.7262} {\path{arXiv:1411.7262}},
  \href {https://doi.org/10.1088/0004-637X/805/1/10}
  {\path{doi:10.1088/0004-637X/805/1/10}}.

\bibitem{LIGOTN}
LIGO, Response of ligo to gravitational waves at high frequencies and in the
  vicinity of the fsr (37.5 khz), Technical Note LIGO-T0060237-00 (2005).

\bibitem{Marolf:2003bb}
D.~Marolf, {On the quantum width of a black hole horizon}, Springer Proc. Phys.
  98 (2005) 99--112.
\newblock \href {http://arxiv.org/abs/hep-th/0312059}
  {\path{arXiv:hep-th/0312059}}, \href
  {https://doi.org/10.1007/3-540-26798-0_9}
  {\path{doi:10.1007/3-540-26798-0_9}}.

\bibitem{CHM}
H.~Casini, M.~Huerta, R.~C. Myers, {Towards a derivation of holographic
  entanglement entropy}, JHEP 05 (2011) 036.
\newblock \href {http://arxiv.org/abs/1102.0440} {\path{arXiv:1102.0440}},
  \href {https://doi.org/10.1007/JHEP05(2011)036}
  {\path{doi:10.1007/JHEP05(2011)036}}.

\bibitem{VZ2}
E.~Verlinde, K.~M. Zurek, {Spacetime Fluctuations in AdS/CFT}, JHEP 04 (2020)
  209.
\newblock \href {http://arxiv.org/abs/1911.02018} {\path{arXiv:1911.02018}},
  \href {https://doi.org/10.1007/JHEP04(2020)209}
  {\path{doi:10.1007/JHEP04(2020)209}}.

\bibitem{Chou:2017zpk}
A.~Chou, et~al., {Interferometric Constraints on Quantum Geometrical Shear
  Noise Correlations}, Class. Quant. Grav. 34~(16) (2017) 165005.
\newblock \href {http://arxiv.org/abs/1703.08503} {\path{arXiv:1703.08503}},
  \href {https://doi.org/10.1088/1361-6382/aa7bd3}
  {\path{doi:10.1088/1361-6382/aa7bd3}}.

\bibitem{Christiansen:2009bz}
W.~A. Christiansen, Y.~J. Ng, D.~J.~E. Floyd, E.~S. Perlman, {Limits on
  Spacetime Foam}, Phys. Rev. D83 (2011) 084003.
\newblock \href {http://arxiv.org/abs/0912.0535} {\path{arXiv:0912.0535}},
  \href {https://doi.org/10.1103/PhysRevD.83.084003}
  {\path{doi:10.1103/PhysRevD.83.084003}}.

\bibitem{He:2014laa}
T.~He, V.~Lysov, P.~Mitra, A.~Strominger, {BMS supertranslations and Weinberg?s
  soft graviton theorem}, JHEP 05 (2015) 151.
\newblock \href {http://arxiv.org/abs/1401.7026} {\path{arXiv:1401.7026}},
  \href {https://doi.org/10.1007/JHEP05(2015)151}
  {\path{doi:10.1007/JHEP05(2015)151}}.

\bibitem{Strominger:2014pwa}
A.~Strominger, A.~Zhiboedov, {Gravitational Memory, BMS Supertranslations and
  Soft Theorems}, JHEP 01 (2016) 086.
\newblock \href {http://arxiv.org/abs/1411.5745} {\path{arXiv:1411.5745}},
  \href {https://doi.org/10.1007/JHEP01(2016)086}
  {\path{doi:10.1007/JHEP01(2016)086}}.

\bibitem{Hawking:2016msc}
S.~W. Hawking, M.~J. Perry, A.~Strominger, {Soft Hair on Black Holes}, Phys.
  Rev. Lett. 116~(23) (2016) 231301.
\newblock \href {http://arxiv.org/abs/1601.00921} {\path{arXiv:1601.00921}},
  \href {https://doi.org/10.1103/PhysRevLett.116.231301}
  {\path{doi:10.1103/PhysRevLett.116.231301}}.

\end{thebibliography}

\end{document}